\documentclass[11pt]{article}
\newtheorem{Theorem}{Theorem}

\newtheorem{Lemma}[Theorem]{Lemma}

\newcommand{\ul}{u^{*}}

\newcommand{\ts}{T^{*}}

\newcommand{\umm}{u_{\mbox{max}}}
\newcommand{\xs}{x^{*}}
\newcommand{\ub}{\bar{u}}
\newcommand{\yb}{\bar{y}}

\newcommand{\pr}{\mbox{Prob}}
\newcommand{\qm}{q^{-}}
\newcommand{\qp}{q^{+}}
\begin{titlepage}
\end{titlepage}
\input{tcilatex}

\begin{document}

\title{On the large time asymptotics of decaying Burgers turbulence.}
\author{Roger Tribe\thanks{%
tribe@maths.warwick.ac.uk}~ and Oleg Zaboronski\thanks{%
olegz@maths.warwick.ac.uk}\\
Mathematics Institute, University of Warwick}
\maketitle

\begin{abstract}
The decay of Burgers turbulence with compactly supported Gaussian "white
noise" initial conditions is studied in the limit of vanishing viscosity
and large time. Probability distribution functions and moments for both
velocities and velocity differences are computed exactly, together with the
"time-like" structure functions $T_{n} (t, \tau) \equiv \langle (u(t+\tau)
-u(t))^n \rangle$.

The analysis of the answers reveals both well known features of Burgers
turbulence, such as the presence of dissipative anomaly, the extreme
anomalous scaling of the velocity structure functions and self similarity of
the statistics of the velocity field, and new features such as the extreme
anomalous scaling of the "time-like" structure functions and the
non-existence of a global inertial scale due to multiscaling of the
Burgers velocity field.

We also observe that all the results can be recovered using the one point
probability distribution function of the shock strength and discuss the
implications of this fact for Burgers turbulence in general.
\end{abstract}

\section{Introduction.}

The study of decaying Burgers turbulence (DBT) is largely motivated by the
the observation that this is a system which falls into the phenomenological
class of turbulent systems which can be treated in principle by means of
Kolmogorov theory. Yet the answers which can be derived analytically for
Burgers turbulence are in the sharp contradiction to the predictions of
Kolmogorov theory. The understanding of the reasons for such a discrepancy
and their relevance for the general theory of turbulence is one of the major
aims of the study of Burgers turbulence.

The history of the subject (see e.g. \cite{Burgers}, \cite{Kida}, \cite
{parker},  \cite{sinai}, \cite{bob}, \cite{Polyakov}, \cite{migdal}, \cite{av1},
\cite{av2}, \cite{av3}, \cite{aurell}, \cite{grisha}, 
\cite{grisha1}, \cite{Gurbatov1}, \cite{khanin1}%
, \cite{khanin}, \cite{truman},  \cite{Truman}, \cite{Gaw}, \cite{bob1}, 
\cite{av}, \cite{weinan}, \cite{FrischBec}; see \cite{Gurbatov} for a review)
shows however that the problem is hard, so hard in fact that it has a
tendency to become self justifying, getting more and more alienated from the
main body of turbulent research. However, until recently
there existed no model
of Burgers turbulence which can be used as a testing ground for general
phenomenological theories of turbulence on one hand and admits a complete
and simple analytical treatment on the other. 

In present paper we introduce and analyse such a model. Namely, we study the
decay of Burgers turbulence with compactly supported Gaussian "white noise"
initial conditions. In physical terms the turbulence in our model is excited
by an initial disturbance localized at a fixed scale much less than the size
of reservoir and which can occur with equal probability around any point of
the reservoir. Note that DBT driven by "white noise" plays a special role
for the theory of DBT in general. The reason is that integral scale of
turbulence in this problem is not imposed by initial conditions but rather
is generated by time evolution. Thus, the answers one obtains for "white
noise" DBT are in some sense universal. Consider for example DBT driven by
Gaussian initial conditions characterized by the two point function $\chi (r)
$ which is approximately constant for $r<<R$ and goes to $0$ exponentially
fast for $r>>R$. Then the statistics of the velocity field in this model at
scales much larger than $R$ and much less than the integral scale is
asymptotically equivalent, in the limit as $\nu \rightarrow 0, t \rightarrow
\infty$, to that of "white noise" DBT. Likewise, compactly supported "white
noise" DBT defines a universality class of models of DBT driven by compactly
supported Gaussian initial conditions.

The choice of a simple initial condition and the choice to look for answers
only in the vanishing viscosity and large time limits lead to a model that
is exactly solvable. Explicit asymptotics can be obtained for statistics
that are hard to estimate in more general models. The main reason for the
exact solvability of our model is the fact that the statistics of the
velocity field in the case of compactly supported initial conditions are
dominated in the limit $\nu \rightarrow 0, t \rightarrow \infty$ by
two shock configurations, the statistics of which is easily computable as
functionals of white noise. 

We would like to stress that our model is in a different universality
class than the original Burgers model in which turbulence is
initiated by white noise initial conditions but no restriction
of compactness is imposed: a solution to Burgers equation corresponding
to an initial condition supported on a whole line will generically
contain infinitely many shocks at any moment of time, not just two
as in our case. Accordingly, the large time statistics of the velocity
field in our case is very different from that in Burgers' model.
For instance, energy density decays as $t^{-1/2}$  in our case (see section
3.1)
and as $t^{-2/3}$ in Burgers', \cite{Burgers}.

The paper is organized as follows. In section 2 we give a precise statement
of the problem, construct a large time limit of the solution to the inviscid
Burgers equation corresponding to compactly supported initial conditions and
formulate the main statements about the statistics of these solutions. In
section 3 we obtain asymptotics for a variety of statistics: the moments of
velocity field; the probability distribution function of velocities; the
velocity structure functions; the probability distribution function of
velocity differences; time-like velocity structure functions. In section 4
the analysis of these results is given. In particular, the validity of
one shock approximation and multiscaling in the problem are discussed.

\section{The limiting velocity field.}

Consider the following initial value problem connected to the Burgers
equation: 
\begin{eqnarray}
\frac{\partial u}{\partial t}(x,t)+u(x,t) \frac{\partial u}{\partial x}%
(x,t)=\nu \frac{\partial ^2 u}{\partial x^2}(x,t),~x \in \mathbf{R}, ~t>0 ,
\label{BE} \\
u(x,0)=u_{0} (x),  \label{IC}
\end{eqnarray}
where $u_{0} (x)$ is a bounded function which is compactly supported in the
interval $[ x_{0}-l, x_{0} +l ]$. Here $l$ is a fixed positive constant and $%
x_{0}$ is a random variable uniformly distributed in the interval $[ -L , L ]
$. The fixed positive constant $L$ plays a role of normalization length.
Conditional on $x_0$ the initial velocity $u_{0} (x)$ will be a white noise
over the interval $[ x_{0}-l, x_{0} +l ]$, so that it has a formal density 
\begin{eqnarray}
P ( u_{0}|x_{0} ) = \frac{1}{Z} e^{-\frac{1}{2J}\int_{x_0 -l} ^{x_0 +l }
u_{0}^2 (x) dx},  \label{prob}
\end{eqnarray}
where $Z$ is a normalization constant chosen in such a way that, formally, 
\begin{eqnarray*}
\int_{-L}^{L} \frac{dx_{0}}{2L} \int P(u_{0}|x_{0}) D(u_{0}) = 1.
\end{eqnarray*}
$J$, the Gaussian variance, is a positive constant which plays a role of
Loitsansky integral for the problem at hand.

Since we have a compact initial condition the distribution of the velocities 
$u_t(x)$ are not translation invariant. The role of $x_0$ is to randomise
the location of the initial disturbance uniformly over the interval $[-L,L]$%
. The values of $u_t(x)$ at a fixed $x$ will then typically be non-zero only
with probability $O(L^{-1})$. We take the limit as $L \rightarrow \infty$
and all the answers concerning the statistics of the velocity field will be
expressed in the form of the leading term in an asymptotic expansion in $%
L^{-1}$. This has the advantage that the answers are then translation
invariant and we are free to consider statistics centered at the origin.

In what follows we will compute asymptotics of the following statistics: the
moments of velocity distribution $M_{n} = \langle u^{n} (x,t) \rangle$; the
velocity structure functions $S_{n} (y) =\langle (u(x+y,t)-u(x,t))^n
\rangle$, the
probability distribution function of the velocity field $P(u) = 
\langle \theta
(u-u(x,t))\rangle$;
the probability distribution function of velocity differences $%
P(u,y) = \langle\theta (u- u(x+y,t) + u(x,t)) \rangle$; 
and the "time-like" velocity
structure functions, $T_{n} (\tau, t)=\langle (u(x,t+\tau)-u(x,t))^n \rangle$%
. Here $\theta(z)=\chi (z \geq 0)$ is the Heavyside function 
and $\langle \ldots
\rangle$ denotes the average w.r.t. to the random initial velocity field $%
u_{0} (x)$.

The solution of the initial value problem (\ref{BE}), (\ref{IC}) for $\nu >0$
via the Cole-Hopf transformation and the evaluation of the limit as $\nu
\rightarrow 0$ for fixed $t>0$ are well known. We refer the reader to \cite
{Hopf} and \cite{Burgers} for a detailed description and give here a quick
summary, sufficient for our needs. The vanishing viscosity solution can be
obtained by plotting a chain of parabolic arcs such that each is touching
the graph of the function $-q(x)=-\int_{-\infty}^{x} u_{0} (y) dy$ at two
points exactly. The i-th parabolic arc is given by a graph of the function $%
\Phi _{i} (x,t) =\Phi _{i} +\frac{(x-x_{i})^2}{2t}$. As time grows the
parabolic arcs flatten out and merge, and there exists a time $T^{*}$ such
that for any $t>T^{*}$ there are generically only two arcs left. The
velocity field associated with such a configuration is then given by 
\begin{eqnarray}
u^{*} (x,t)= U(x_0+x^*,x,t,P,Q) \equiv \frac{(x-x_{0}-x^{*})}{t}
\chi_{[(x_{0}+x^{*}- \sqrt{-2Qt}, x_{0}+x^{*} + \sqrt{2(P-Q)t}]} (x),
\label{u}
\end{eqnarray}
where $\chi _{I}$ is an indicator function of the interval $I$, $P=
q(+\infty)$ is a momentum corresponding to a given $u_{0}$, $Q=\min_{x} q(x)$
is a global minimum of $q(x)$ and $x_{0}+x^{*} \in [x_{0} -l, x_{0} +l]$ is
the point where this minimum is achieved. (Such a point exists and is unique
almost surely as $q(x)$ is continuous and the global minimum is almost
surely unique.) The limiting solution (\ref{u}) was originally constructed
in \cite{Hopf}.

The time $T^*$ at which the limiting velocity field $u^{*}$ is attained
depends on the random initial condition $u_0$ but it will be shown that the
statistics of the velocity field is well approximated at large times by the
statistics of the limiting velocity field $u^{*}$. The latter is determined
in turn by the joint distribution of the momentum $P$ and the global
minimum $Q$. Indeed although the expression for $u^{*}$ depends explicitly
on $(P,Q,x^{*})$, the dependence on $x^*$ doesn't influence the statistics
of $u^{*}$ in the limit $L \rightarrow \infty$, where the translational
invariance is restored. We delegate the detailed discussion of this point
to the next section.

The choice of white noise as an initial distribution leads to the
distribution of the pair $(P,Q)$ being exactly calculable. Indeed it is a
well known consequence of the 'reflection principle' for Brownian paths \cite
{Revuz}. Since it is key to all our asymptotics we include a quick
derivation of the joint density function $\rho (P,Q)$. We start with a
computation of the probability distribution function of momentum $\rho (P)$.
Writing $\delta$ for the delta function at zero, we have by definition 
\begin{eqnarray*}
\rho (P) = \left\langle \delta \bigg( P - \int_{-\infty}^{\infty}dx \, 
u_{0} (x) \bigg) \right\rangle=
\int_{-\infty}^{\infty} \frac{d \lambda}{2 \pi} e^{i \lambda P} \left\langle
e^{-i \lambda \int_{-\infty}^{\infty} dx u_{0} (x) }\right\rangle.
\end{eqnarray*}
Using (\ref{prob}) this functional integral is Gaussian and can be simply
computed to give 
\begin{eqnarray*}
\left\langle e^{-i \lambda \int_{-\infty}^{\infty} dx u_{0} (x)
} \right\rangle=e^{-lJ\lambda ^2}.
\end{eqnarray*}
The integral over $\lambda$ is a Gaussian integral and we conclude that the
distribution of $P$ is also Gaussian, as could have been guessed from the
very beginning, and given by 
\begin{eqnarray}
\rho (P) = \frac{e^{-(\frac{P}{P_{0}})^2}}{\sqrt{\pi} P_{0}}, \qquad 
\mbox{where 
$ P_{0} = 2\sqrt{lJ} .$}  \label{prP}
\end{eqnarray}
The joint probability distribution function can now be computed as follows.
Fix $q,p$ satisfying $q < 0, q < p$. Let $x^{\prime}$ be the first value of $%
x$ for which $q(x) =q$. Define $q^{\prime}(x)$ to equal $q(x) $ for $x \leq
x^{\prime}$ and to equal the reflection of $q(x) $ in the horizontal line $%
y=q$ for $x \geq x^{\prime}$. Then if $Q^{\prime}= \min_x q^{\prime}(x)$ and 
$P^{\prime}= q^{\prime}(\infty)$ the reflection principle (see \cite{Revuz}),
which exploits the white noise nature of $u_0$, states that $%
Q^{\prime},P^{\prime}$ have the same distribution as $Q,P$. Then 
\begin{eqnarray*}
\mbox{Prob} (Q \leq q, P \geq p) &=& \mbox{Prob} ( Q^{\prime}
 \leq q, P^{\prime}\leq
2q-p) \\
&=& \mbox{Prob} (P^{\prime}\leq 2q-p) \\
&=& \int_{-\infty}^{2q-p} \frac{dz}{\sqrt{\pi} P_{0}} e^{-(\frac{z}{P_{0}}%
)^2}.
\end{eqnarray*}
Differentiating in $p$ and $q$ we conclude that 
\begin{eqnarray}
\rho (p,q) = \frac{4(p-2q)}{\sqrt{\pi} P_{0}^3} e^{-(\frac{p-2q}{P_{0}})^2}, %
\mbox{ if } ~ q \leq min\{ 0,p \} ,  \label{prPQ}
\end{eqnarray}
and is zero for all other values of $p$ and $q$.

With the help of (\ref{prPQ}) we are able to average functionals $F[u^*(t)]=
F [ u^{*} (x_i,t): i=1,2,\ldots]$ with respect to the initial distribution.
If however we are interested in the statistics of $u (x,t)$ at zero
viscosity and large times there is still a question: is it true that in this
limit $\langle F[u(t)] \rangle \sim \langle F[u^{*} (t)] \rangle$, or even
at large times are there statistically many initials conditions such that
corresponding velocity profiles haven't converged to the limiting ones? It
so happens that the first alternative prevails. The detailed proofs of this
fact for relevant functionals are carried out in the next section and in the
appendix and are based on the following estimate on the time $T^{*}$ of
convergence to the limiting profile: 
\begin{eqnarray}
\mbox{Prob} (T^{*} > t) \leq C \left( \frac{t_{c}}{t} \right) ^{\frac{1}{2}}
\label{ts}
\end{eqnarray}
where $t_{c}=\sqrt{\frac{l^3}{J}}$ and $C$ is a positive number. The proof
of this estimate is fairly complicated and is allocated to
the appendix. However the result itself is so important for the
validity of conclusions of our paper that we decided to present here a
convincing and very simple heuristic derivation of it.

By definition, $\pr (\ts < t \mid P, Q, \xs) = \pr (q \leq \Phi _{t}
\mid P, Q, \xs)$, where $\Phi _{t}$ coincides for $x<\xs$
with with 
parabolic arc $\Phi _{1,t}$
passing through the point $(\xs, -Q)$ and touching the line
$y=0$ and with parabolic arc $\Phi _{1,t}$
passing through the point $(\xs, -Q)$
and touching the line $y=-P$ for $x>\xs$ (we used the translation
invariance of the random variable $\ts$ to set $x_{0}=0$. Consequently,
$\xs \in [-l,l]$).

It is convenient to think of
a Brownian walk $q(x)$ passing through $(\xs, -Q)$ as a collection of
two independent walks $\qp (x)$
and $\qm(x)$
starting at this point and moving in the opposite
directions in "time" $x$. Therefore, 
\begin{eqnarray}
\pr (\ts < t \mid P,Q, \xs) = 
\pr (\qm < \Phi _{1, t} \mid Q, \xs) \cdot
\pr (\qp <\Phi _{2,t} \mid P,Q, \xs) .
\label{mastpr}
\end{eqnarray}
To estimate, say, $\pr (\qm < \Phi _{1, t} \mid Q, \xs)$ below
we note that $\pr (\qm < \Phi _{1, t} \mid Q, \xs) \geq
\pr (\qm < -Q + \theta \cdot (x-\xs) \mid Q, \xs)$, 
where $y=-Q + \theta \cdot (x-\xs)$ is an equation for 
the line tangent to
the parabola $\Phi _{1,t}$ at the point $(\xs, -Q)$;
$\theta = \sqrt{\frac{-2Q}{t}}$. 
Hence,
\begin{eqnarray}
\pr (\qm < \Phi _{1, ~t} \mid Q, \xs) \geq \nonumber \\
\geq
\lim_{\epsilon \to +0}
\frac{\int\limits_{q(-l) =0}^{q(\xs)=-Q-\epsilon} Dq ~ \Theta 
\bigg[ q < -Q + \theta \cdot (x-\xs) \bigg] ~ 
e^{-\frac{1}{2J}\int_{-l}^{\xs} \dot{q} ^2 dx}}
{\int\limits_{q(-l) =0}^{q(\xs)=-Q-\epsilon} Dq ~ \Theta 
\bigg[ q < -Q \bigg] ~ e^{-\frac{1}{2J}\int_{-l}^{\xs} \dot{q} ^2 dx}}
\Theta \bigg( 0<-Q-\theta \cdot (l+\xs) \bigg) ,
\label{funct}
\end{eqnarray}
where $\Theta [\ldots ]$ is a functional step function,
$\Theta (\ldots )$ - a usual one. 

The functional integral in the numerator of (\ref{funct})
can be transformed into an integral over all pathes satisfying
$q(x) \geq 0$ by a change of variables $q(x) \rightarrow
q(x)-Q+\theta (x-\xs )$. (A counter part of this 
transformation in quantum
mechanics is a Galilean transformation.) Now the functional
integrals in both numerator and denumerator of (\ref{funct})
can be expressed in terms of Green's function of heat equation
$\dot{q}=\frac{J}{2} q^{\prime \prime}$ on half a line, i.e.
the antisymmetrization of Green's function of the same equation
on the whole line. A simple computation shows then that
\begin{eqnarray}
\pr (\qm < \Phi _{1, t} \mid Q, \xs)  
\geq \left(1 -\sqrt{\frac{2l^2}{-Qt}} \right) \cdot
\theta \left( 1-\bigg( \frac{2l^2}{-Qt} \bigg) \right) .
\label{est11}
\end{eqnarray}
Similar estimate 
holds for $\pr (\qp < \Phi _{2, t} \mid P,Q, \xs)$ if one
replaces $-Q$ with $P-Q$ in the r. h. s. of (\ref{est11}).

Substituting these two estimates into (\ref{mastpr}) and
integrating both sides of the resulting inequality w. r. t.
$P,Q$ using (\ref{prPQ}) we find that $\pr (\ts < t) \geq 1-Const
\sqrt{\frac{l^2}{tP_{0}}}$, which is equivalent to the estimate
(\ref{ts}) for $\pr (\ts >t) = 1- \pr (\ts < t)$.  

\section{The statistics of the velocity field in the $\nu \rightarrow 0$, $t
\rightarrow \infty$ limit.}

\subsection{Moments of the velocity distribution.}

\label{3.1}

The aim of the present section is to compute the large $t$-limit of moments
of the velocity distribution 
\begin{eqnarray}
M_{n} (t) = \left\langle u^{n} (0,t) \right\rangle, ~n=1,2,\dots  \label{M}
\end{eqnarray}
Odd order moments vanish identically due to the symmetry: both Burgers
equation and the initial distribution are invariant with respect to the
transformation $u \rightarrow -u$, $x \rightarrow -x$. On the other hand, $%
M_{2k+1} \rightarrow - M_{2k+1}$ under this transformation, which implies
that $M_{2k+1}(t) \equiv 0 $ for $k=1,2,\ldots$ We concentrate therefore on
the computation of the moments of even order and assume everywhere below
that $n$ is even. We may write, using the fact that $u(x,t) = u^{*} (x,t)$
for $t > T^{*}$, 
\begin{eqnarray}
M_{n}(t) = \left\langle u^{*n} (0,t) \right\rangle + R_{n} (t),  \label{M*}
\end{eqnarray}
where 
\begin{eqnarray}
R_{n} (t) = \left\langle \bigg( u^{n} (0,t) - u^{* n} (0,t) \bigg) 
\theta (T^* -t) \right\rangle
\label{Rn}
\end{eqnarray}
is an error term to be estimated.

The first term in the right hand side of (\ref{M*}) can be written in the
following form: 
\begin{eqnarray}
\left\langle u^{*n} (0,t) \right\rangle = \int dp dq \; \rho (p,q) \int_{-L}^{L} \frac{%
dx_{0}}{2L} \; U^n(x_0,0,t,p,q) +r_{n} (t),  \label{u*n}
\end{eqnarray}
where 
\begin{eqnarray}
r_{n} (t)=\int_{-l}^{l} dx^* \int dp dq \; \rho (p, q, x^*) \bigg( \int_{-L
+x^{*}}^{-L} + \int_{L}^{L+x^*} \bigg) \frac{dx_{0}}{2L} \; U^n(x_0,0,t,p,q)
\label{rn}
\end{eqnarray}
is an error term appearing due to neglecting $x^{*}$ in comparison to $L$
and $\rho (p, q, x^*)$ is a joint probability density of $P, Q$ and $x^*$.
It is shown in the appendix that the error term $r_{n} (t)$ does not affect
the asymptotics as $\nu \rightarrow 0, L \rightarrow \infty , t \rightarrow
\infty$. Informally this fact can be explained by noticing that the
integrand in (\ref{rn}) is non-zero only for velocity profiles which are
"stretched" over the interval of length $L$ and thus are exponentially
improbable.

The remaining integral on the right hand side of (\ref{u*n}) can be
evaluated exactly using the explicit expressions (\ref{u}) and (\ref{prPQ}%
) leading to  the following result: 
\begin{eqnarray}
\left\langle u^{*n} (0,t) \right\rangle \sim \frac{\Gamma ((n+3)/4)}{\sqrt{\pi} (n+1)} 
\frac{L(t)}{L} U(t)^{n}  \label{u*n2}
\end{eqnarray}
where 
\begin{eqnarray}
L(t)=\sqrt{2P_{0} t}, ~U(t) = \frac{L(t)}{t}  \label{scale}
\end{eqnarray}
are parameters, with dimensions length and velocity, which should be
interpreted as the scale of turbulence and turbulent velocity
correspondingly. Here we write the symbol $\sim$ to mean asymptotic
equivalence in the limit as $L \rightarrow \infty$ and then $t \rightarrow
\infty$.

Another computation presented in the appendix leads to the following
estimate of the error term $R_{n} (t)$ from (\ref{M*}): 
\begin{eqnarray}
|R_{n} (t)| \leq C_{n} \frac{L(t)}{L} U(t)^n 
\left( \frac{t_{c}}{t} \right) ^{1/4},
\label{Rnest}
\end{eqnarray}
where $t_{c} = \sqrt{l^3/J}$ is a constant having a dimension of time, $%
C_{n} $ is a positive constant. Comparing (\ref{Rnest}) with (\ref{u*n2}) we
see that for $t>>t_{c}$, $|R_{n} (t)| << \langle u^{*n} (0,t) \rangle$,
which permits us to conclude that 
\begin{eqnarray}
M_{2k} (t) \sim \frac{\Gamma (k/2+3/4)}{\sqrt{\pi} (2k+1)} \frac{L(t)}{L}
U(t)^{2k}, \qquad k=1,2,\ldots  \label{moments}
\end{eqnarray}

It is important to stress however that coefficient $C_{n}$ from (\ref{Rnest})
grows faster with $n$ 
than the number factor in the r. h. s. of (\ref{moments}). Thus it takes
a long time for a moment of high order to converge to 
the limiting value (\ref{moments}).

It follows from (\ref{moments}) that the energy density $E(t) \equiv \frac{1%
}{2} M_{2} (t)$ decays like $t^{-1/2}$ as $t \rightarrow \infty$.  
This is the result to be expected: Dissipation of energy
occurs in Burgers turbulence due to shock collisions and at each separate
shock. The energy of a separate shock decays as $t^{-1/2}$ and due to the 
absence of shock collisions in the limiting profile (\ref{u}), this also
gives the law of decay of total energy density. This argument
is due to J. M. Burgers, see \cite{Burgers}.

We will also see below that the statistics of
the velocity field in our model is self-similar with the scales of length
and velocity given by (\ref{scale}). These scales depend on time exactly as
their counterparts in Kida's model. The statistics of the velocity field in
our case are however different from that of  Kida\footnote{%
There exists no complete solution of Kida's model. Yet the answers which can
be obtained within Kida's model are different from their counterparts in our
model.}. Thus we conclude that the self-similarity alone does not determine
the large time asymptotics of the statistics of velocity field in DBT. Note
also that $E(t)$ decays in time, showing the presence of a dissipation
anomaly in the model: the rate of energy dissipation does not vanish but
converges to a finite non-zero limit when the viscosity $\nu$ approaches
zero.

\subsection{The probability distribution function of velocities.}

In this section we will concern ourselves with computing the probability
distribution function (PDF) of velocities given by 
\begin{eqnarray}
P(u,t) \equiv \mbox{Prob} (u(0,t)>u)=\left\langle \theta (u(0,t)-u) \right\rangle .
\label{pru}
\end{eqnarray}
Reasoning exactly as in the previous section we find that 
\begin{eqnarray}
P(u,t)=\left\langle \theta (u^{*} (0,t)-u) \right\rangle +R(u,t),  \label{pru2}
\end{eqnarray}
where 
\begin{eqnarray}
R(u,t)= \left\langle \left(\theta \bigg(u(0,t)-u\bigg) -\theta \bigg( u^{*}
(0,t)-u \bigg) \right) \theta (t^{*} -t) \right\rangle  \label{R}
\end{eqnarray}
is an error due to the replacement $u \rightarrow u^{*}$; 
\begin{eqnarray}
\left\langle \theta (u^{*} (0,t)-u) \right\rangle = \nonumber \\
=\theta (-u) +\int dpdq \, \rho (p,q)
\int_{-L}^{L} \frac{dx_{0}}{2L} \bigg( \theta (U(x_0,0,t,p,q)-u) - \theta
(-u) \bigg) +r(u,t),  \label{prus}
\end{eqnarray}
where $r(u,t)$ is an error due neglecting $x^{*}$ in comparison with $L$: 
\begin{eqnarray}
r(u,t)= \nonumber \\
=\int_{-l}^{l}d x^{*} \int dpdq\, \rho (p,q,x^{*} )
\bigg( \int_{-L+x^{*}}^{-L}+\int_{L}^{L+x^{*}} \bigg) 
\frac{dx_{0}}{2L} \left(\theta
(U(x_0,0,t,p,q)-u) - \theta (-u) \right).  \label{r}
\end{eqnarray}
The reason that the term $\theta(-u)$ is added and subtracted is that, due
to the averaging of the position of the initial condition over the block $%
[-L,L]$, the velocity is typically zero and so the PDF is an $O(L^{-1})$
perturbation to $\theta(-u)$.

An estimate of $r(u,t)$ similar to that of the term $r_n(t)$ in section (\ref
{3.1}) shows that $r(u,t)$ does not affect the final asymptotics. An exact
calculation using the known density $\rho(p,q)$ for the other terms on the
right hand side of (\ref{prus}) leads to 
\begin{eqnarray}
\left\langle \theta (u^{*} (0,t)-u) \right\rangle \sim \theta (-\bar{u}) +\frac{L(t)}{L}
\int_{\bar{u}^2}^{\infty} \frac{d \alpha}{\sqrt{\pi}} e^{-\alpha ^2} 
\bigg(\sqrt{%
\alpha} - |\bar{u}| \bigg) \mbox{sgn}(\bar{u} )  \label{prus2}
\end{eqnarray}
where $\bar{u} = u/U(t)$. A computation performed in the appendix shows that 
\begin{eqnarray}
| R(u,t) | \leq C \frac{L(t)}{L} (\frac{t_{c}}{t})^{1/4}  \label{Rest}
\end{eqnarray}
where $C$ is a positive constant. Comparing (\ref{Rest}) with (\ref{prus2})
we see that for $t >>t_{c}$ we have $\langle \theta (u^{*} (0,t)-u) \rangle
>> |R(u,t)|$, with the last inequality being pointwise in $\bar{u}$ rather
than uniform. We conclude that 
\begin{eqnarray}
P(U(t) \ub,t) \sim \theta (-\ub) + \frac{L(t)}{L} \int_{\ub ^2}^{\infty}
\frac{d
\alpha}{\sqrt{\pi}} e^{-\alpha ^2} 
\bigg( \sqrt{\alpha} - |\ub| \bigg) \mbox{sgn}(\ub )
\label{pru3}
\end{eqnarray}
If in particular $|\ub| \rightarrow \infty$ this simplifies to 
\begin{eqnarray}
P(U(t) \ub, t) \sim \theta (-\ub) + \frac{1}{8\sqrt{\pi}} 
\mbox{sgn}(\ub) \frac{%
L(t)}{L} \frac{e^{-\ub ^{4}}}{\ub ^5}.  \label{pruas}
\end{eqnarray}

Note that the answer (\ref{pru3}) for $P(u,t)$ is self-similar with $U(t)$
playing the role of the integral velocity scale. Note also that the form of $%
P(u,t)$ is not Gaussian. This confirms the non-triviality of our model: the
output (the strongly non-Gaussian statistics of the velocity field in the
limit of small viscosity and large time) is not the same as the input (a
trivial Gaussian distribution of the initial velocity field). This
non-triviality will be re-emphasised in the consequent sections where it will
be shown that the limiting statistics of the velocity field is intermittent.

Finally we would like to make the following technical comment. Of course,
the moments of the distribution (\ref{pru3}) are exactly those given by (\ref
{moments}). We could therefore try to compute the distribution (\ref{pru})
first and then argue that the moments of this asymptotic distribution
coincide with the asymptotics of the moments of the actual distribution.
Unfortunately the analysis of error terms within this approach becomes very
involved. For this reason we have two separate computations, the asymptotics
of the moments of the velocity distribution and the asymptotics of the
velocity distribution itself.

\subsection{Velocity structure functions.}

Now we will turn to the two-point statistics of the velocity field and
compute asymptotics for the velocity structure functions given by 
\begin{eqnarray}
S_{n} (y,t)=\left\langle \bigg( u(y,t)-u(0,t) \bigg) ^n
\right\rangle  \quad n=1,2,\ldots
\end{eqnarray}
We find as in the previous subsections that 
\begin{eqnarray}
S_{n} (y,t) = \left\langle \bigg( u^{*} (y,t)- u^{*} 
(0,t) \bigg) ^n \right\rangle  
+ R_{n} (y,t),  \label{corr}
\end{eqnarray}
where $R_{n} (y,t) $ accounts for the error due to the replacement of $u$
with $u^{*}$. As shown in the appendix this error can be estimated as
follows:  for $t$ such that $L(t) \geq y$ 
\begin{eqnarray}
| R_{n} (y,t) | \leq C_{n} U^{n} (t) \frac{y}{L} 
\bigg( \frac{t_{c}}{t} \bigg) ^{1/4}.
\label{est1}
\end{eqnarray}
We express the first term in (\ref{corr}) as 
\begin{eqnarray*}
\left\langle \bigg( u^{*} (y,t)- u^{*} (0,t)
\bigg) ^n \right\rangle  = \\
=\int_{-L}^{L} \frac{dx_{0}}{2L} 
\int dp dq
\, \rho (p,q) \bigg( U(x_0,y,t,p,q) - U(x_0,0,t,p,q) \bigg) ^n + r_{n} (y,t)
\end{eqnarray*}
where $r_{n} (y,t)$ accounts for an error arising due to neglecting $x^{*}$
in comparison with $L$. Again it can be shown that the term $r_{n} (y,t)$
does not contribute to the asymptotics. Now a direct computation using the
density $\rho(p,q)$ shows, for $y \geq 0$, that 
\begin{eqnarray*}
\left\langle \bigg( u^{*} (y,t)- u^{*} (0,t) \bigg) ^n 
\right\rangle  \; \sim \; 
(-1)^{n} \frac{1}{\sqrt{\pi}}
\Gamma \bigg( \frac{n+2}{4} \bigg) 
\frac{L(t)}{L} U^{n} (t) \bar{y} +O(\bar{y}^2),~
n=2,3,\ldots
\end{eqnarray*}
where $\bar{y} = \frac{y}{L(t)}$. In addition $S_{1} (y,t) \sim 0$, which
confirms the restoration of translation invariance in the large $L$ limit.
Comparing this with (\ref{est1}) we see that the asymptotics of the
velocity structure functions is given, for fixed $\yb \leq 1$, by 
\begin{equation}  \label{sn}
S_n(L(t)\yb ,t) \sim (-1)^{n} \frac{1}{\sqrt{\pi}} \Gamma (\frac{n+2}{4}) 
\frac{L(t)}{L} U^{n} (t) \yb +O(\yb ^2) \quad n=2,3,\ldots
\end{equation}
It has been assumed in our computations that $\yb \geq 0$. 
Extending (\ref{sn})
to negative $y$ by the symmetry $y \rightarrow -y$, $u \rightarrow u$, we
see that $S_{2k}(L(t) \yb,t)$ is proportional to $| \yb|$ 
and $S_{2k+1}(L(t)\yb ,t)$
is proportional to $\yb$ for $k \geq 1$ and $|\yb| <<1$.

Thus the velocity structure functions of the problem exhibit in the inertial
range the extreme anomalous (non-Kolmogorov) scaling which is typical for
Burgers turbulence in general and is due to the presence of shocks in the
limiting velocity profile. The Burgers anomalous scaling is well known from
heuristic arguments (see e.g. \cite{Frisch}, \cite{Grisha}, \cite{Gaw} ). In
our case however it has been derived as a part of the complete solution of
the problem.

\subsection{The probability distribution function of velocity differences.}

Here we will compute the PDF for velocity differences 
\begin{eqnarray}
P(u,y,t) = \mbox{Prob} \bigg(u > \Delta u(y,t)\bigg) =
\left\langle \theta \bigg( u-\Delta u(y,t) \bigg) \right\rangle,
\label{prd}
\end{eqnarray}
where $\Delta u (y,t) = u(y/2,t)-u(-y/2,t)$ and $y \geq 0$. Definition (\ref
{prd}) is tailored for the study of negative velocity differences and we
consider only the case $u<0$. Negative differences are the interesting case
since they occur when the velocities are evaluated either side of a shock. 
A lengthy but straightforward computation shows that for fixed $\ub<0$, $\yb>0$
\begin{eqnarray}
P(U(t) \ub ,L(t) \yb,t) \; \sim \; \nonumber \\
\; \sim \; 
2\frac{L(t)}{L}\bigg( \int_{\ub ^2}^{(\yb-\ub)^2} \frac{%
d\alpha}{\sqrt{\pi}} e^{-\alpha ^2} \bigg( \sqrt{\alpha} + \ub \bigg) + \yb
\int_{(\yb-\ub)^2}^{\infty} \frac{d\alpha}{\sqrt{\pi}} 
e^{-\alpha ^2}\bigg) +R(\ub,\yb ,t)
\label{prdr}
\end{eqnarray}
where, as shown in the appendix, 
\begin{eqnarray}
| R(\ub,\yb,t) | \leq C \frac{L(t)}{L} 
\frac{\yb}{\yb+|\ub|} \bigg( \frac{t_{c}}{t} \bigg) ^{1/4}.
\label{errtrm}
\end{eqnarray}
Due to the presence of extra factor of $(\frac{t_{c}}{t})^{1/4}$ decaying
with time, $R(u,y,t)$ becomes small compared to the first term in the right
hand side of (\ref{prdr}), given that $\ub, \yb$ fixed.

It is easy to analyse (\ref{prdr}) in the following limiting cases. We
suppose that $\yb << 1$. If $|\ub| <<1$, then 
\begin{eqnarray}
P(U(t)\ub ,L(t)\yb,t) \sim \frac{L(t) \yb}{L}
\left( 1-\frac{\sqrt{\pi}}{2} \yb - 
\frac{\sqrt{\pi}}{2} |\ub|^2 + O(\yb ^2) + O(|\ub|^3) \right).  \label{lim1}
\end{eqnarray}
If $1<< |\ub| << \yb ^{-1/3}$ then 
\begin{eqnarray}
P(U(t) \ub,L(t) \yb ,t) \sim \frac{L(t) \yb}{\sqrt{\pi} L}
\frac{1}{| \ub |^2}
e^{-|\ub |^4} \left( 1+O(\frac{1}{ |\ub |^4}) + 
O(\yb | \ub |^3) \right)  \label{lim2}
\end{eqnarray}
If $| \ub | >> \yb ^{-1/3}$ then 
\begin{eqnarray}
P(U(t) \ub ,L(t) \yb ,t) \sim \frac{L(t)}{4\sqrt{\pi} L} 
\frac{1}{| \ub |^5} e^{-|
\ub |^4} \left( 1+O(\frac{1}{| \ub |^8}) \right)  \label{lim3}
\end{eqnarray}

To summarize, for negative $u$, $P(u,y,t)$ decays algebraically for $| u| <<
U(t)$ and super exponentially for $| u | >> U(t)$. Moreover,
$P(u,y,t) \sim O(y)$ if $1<< |\ub| << \yb ^{-1/3}$ and doesn't
depend on $y$ if $1<< |\ub| << \yb ^{-1/3}$. This information alone 
enables one to conclude that velocity structure functions of sufficiently 
high order exhibit anomalous scaling. In addition we
observe a crossover
between regimes (\ref{lim2}) and (\ref{lim3}). This crossover is actually
responsible for the presence of many scales in description of the
statistics of velocity field and the absence of the universal inertial
range in Burgers turbulence. We refer reader to section 4 for a detailed
discussion of this point.

\subsection{The multi-time statistics of the velocity field.}

The simplicity of our model allows us to compute the correlation between
values of the velocity field at different moments of time. Let 
\begin{eqnarray}
T_{n} (\tau,t) = \left\langle \bigg( u(0,t+\tau)-u(0,t)\bigg) ^n 
\right\rangle  \label{tn}
\end{eqnarray}
be the velocity structure functions corresponding to the same point at space
but different moments of time. We write (\ref{tn}) in the already familiar
form 
\begin{eqnarray}
T_{n} (\tau,t) = \left\langle \bigg( u^{*} (0,t+\tau)-u^{*} (0,t)
\bigg) ^n \right\rangle +R_{n}
(\tau,t)  \label{tns}
\end{eqnarray}
with $R_{n} (\tau)$ accounting for an error due to the replacement of $u$
with $u^{*}$. An estimate in the appendix shows that 
\begin{eqnarray}
| R_{n} (\tau ,t) | \leq C_{n} U^{n} (t) \frac{\tau U(t)}{L} 
\bigg( \frac{t_{c}}{t} \bigg) ^{1/4}.  
\label{rnt}
\end{eqnarray}
The computation of the first term in the right hand side of (\ref{tns}) is
very close to the computation performed in previous sections and leads to 
\begin{eqnarray}
T_{n} (\tau,t) & \sim & (-1)^n \frac{\Gamma (\frac{n+3}{4})}{2 \sqrt{\pi}}
U^{n} (t) \frac{U(t) \tau}{L} +R_{n}(\tau,t)  \nonumber \\
& \sim & (-1)^n \frac{\Gamma (\frac{n+3}{4})}{2 \sqrt{\pi}} U^{n} (t) \frac{%
U(t) \tau}{L}, \quad n=2,3,\ldots  \label{tnf}
\end{eqnarray}
We therefore conclude that the time-like structure functions exhibit in
Burgers turbulence the extreme anomalous scaling in $\tau$ given by $T_{n}
(\tau,t) \sim \tau, n=2,3,\ldots$. Comparing this with the expression (\ref
{sn}) for the space-like structure functions, we see that 
\begin{eqnarray}
S_{n} (y,t) = T_{n} (\tau,t), \qquad n=2,3,\ldots  \label{taylor}
\end{eqnarray}
at $y=C(n) U(t) \tau$, given that $y << L(t) $ and $\tau <<t$. The identity (%
\ref{taylor}) means that "isotropic" Taylor conjecture 
stating the equivalence of the space-like and time-like
statistics in isotropic turbulence at small scales, becomes a theorem for
our model of Burgers turbulence. The similar
observation was also independently
made in \cite{FrischBec} in the context Burgers turbulence generated
by correlated Gaussian initial conditions.

Let us finally note that if one wishes to compare $T_{n} (y,t)$
with $S_{n} (\tau , t)$ at arbitrarily high orders $n$, the condition
of applicability of relation (\ref{taylor}) 
has to be changed to $y << L_{n} (t)$, where $L_{n} (t)$
is correlation length associated
with $n$-th order structure function
introduced in section 4.2. For $n>>1$, $L_{n} (t) \sim
L(t)/n^{3/4}$, see \ref{ln} below.

\subsection{One-shock approximation.}

We wish to show that all of the results obtained in the previous section can
be easily obtained from heuristic arguments given the knowledge of the
probability density of a velocity jump at a shock. In our case the latter is
easy to compute: a simple computation which uses the knowledge of the
limiting velocity profile (\ref{u}) and the density $\rho(p,q)$ gives 
\begin{eqnarray}
\rho (\mu) \equiv \left\langle \delta 
\bigg( \mu - \sqrt{2(P-Q)/t} \bigg) \right\rangle =\frac{4}{%
\sqrt{\pi} U(t)} \bar{\mu} e^{-\bar{\mu} ^4},  \label{sj}
\end{eqnarray}
where $\mu$ is a velocity jump at the (right) shock, $\bar{\mu} = \frac{\mu}{%
U(t)}$. The probability density of the velocity jump at the left shock has
exactly the same form, so we will be referring to (\ref{sj}) as the
probability density of the velocity jump at a shock.

Now let us $assume$: Firstly that the large-$t$ statistics of $u$ are
approximated by that of $\ul$; secondly that a one-shock approximation is
valid, i.e. that one can disregard in the analysis the contributions coming
from configurations with shocks separated by distances much less than the
average separation $L(t)$.

To derive $P(u,y,t) $ for $u<0,~y<<L(t)$ using these assumptions note that $%
u(y,t)-u(0,t)$ can be negative only if there is a shock at some point in $%
[0,y]$. If the right hand shock lies at $x \in [0,y]$ then $u(y,t)-u(0,t) =
-\mu+x/t$. A similar formula holds if the left hand shock lies in $[0,y]$.
So neglecting the contribution from the configurations with 2 shocks inside
the interval $[0,y]$, we see that 
\begin{eqnarray*}
\mbox{Prob} \bigg( u(y,t)-u(0,t) <u \bigg) 
\approx 2 \int_{0}^{y} \frac{dx}{2L} \; %
\mbox{Prob} (\mbox{Size of Jump} > \frac{x}{t} -u ).
\end{eqnarray*}
This can be easily computed using the density of the shock jump (\ref{sj})
giving 
\begin{eqnarray}
\mbox{Prob} \bigg( u(y,t)-u(0,t) < y \bigg) 
\approx \frac{2L(t)}{L} \bigg( \int_{\bar{u} ^2}^{(%
\bar{y}-\bar{u})^2} \frac{d\alpha}{\sqrt{\pi}} e^{-\alpha ^2} (\sqrt{\alpha}
+ \bar{u}) +\bar{y} \int_{(\bar{y}-\bar{u})^2}^{\infty} \frac{d\alpha}{\sqrt{%
\pi}} e^{-\alpha ^2} \bigg) ,
\end{eqnarray}
which coincides with the exact answer $(\ref{prdr})$.

With the knowledge of the PDF of velocity difference we can compute velocity
structure functions, thus moments of velocities, thus the PDF of velocities.
In other words all of the results of the previous section concerning single
time statistics of the velocity field can be obtained using a one-shock
approximation.

Moreover, the $\tau$-dependence of the time-like structure functions (\ref{tnf})
is also entirely due to the one-shock effects: if $n=2,3,...$ and $\tau <<t$%
, then the main contribution to $T_{n} (\tau,t)$ comes from the
configurations with a shock passing through $x=0$ between the moments of
time $t$ and $t+\tau$. A shock with velocity jump $\mu$ travels a distance
approximately $\mu \tau/2$ over the interval $[t,t+\tau]$.  Therefore, 
\begin{eqnarray*}
T_{n} (\tau) & \approx & \left\langle (-\mu)^n \chi \bigg( %
\mbox{Shock passed through $0$ 
during $[t,t+\tau]$} \bigg) \right\rangle \\
& \approx & \langle (-\mu)^n \frac{\mu \tau}{2L} \rangle.
\end{eqnarray*}
Computing this average using the PDF of shock strength (\ref{sj}) we arrive
exactly at (\ref{tnf}), which again shows that one-shock approximation is
asymptotically exact.

These calculations support the following statement about decaying Burgers
turbulence:  all one needs to know in order to describe the statistics of
the velocity field at scales much less than the average distance between
shocks is the $one$-point PDF of shock velocity and strength (or just shock
strength if the correlation functions which we're trying to compute are
Galilean-invariant). Thus the problem is much simpler than one might have
thought: recall for example that exact formulae expressing velocity
correlation functions in terms of the statistics of shocks are such (\cite
{Burgers}, \cite{Kida}) that one seemingly needs to know the $n$-point joint
PDF of shock strengths in order to compute $n$-th order correlation function.

The rigorous proof of the above statement together
with estimates on the errors of one-shock approximation
will make DBT analytically
tractable for a wide class of initial conditions as the great deal is 
known about
the one-point function of shock strength, see e. g. \cite{Burgers},
\cite{Kida}, \cite{sinai}, \cite{av2}, \cite{av3}.

Is there a universal technique for the computation
of the one-point PDF of the shock strength? It has been 
known since Burgers \cite{Burgers}, but never really exploited, that shocks
behave (almost) as a system of sticky particles. One might try therefore to
extract the information about one-point PDF of shock characteristics by
studying the kinetics of this system, for example, by analyzing the
Smoluchowski-Bogoluibov chain of equations for one-point, two-point, $\ldots$
PDF's of shocks.

\subsection{On multiscaling in Burgers turbulence.}

In statistical physics the term "multiscaling", instead of "anomalous
scaling", is used  to stress an inherently multiscale
nature of a system exhibiting anomalous scaling of correlation functions.
Burgers turbulence is no exception. In this section we will show that the
crossover between the tails (\ref{lim2}) and (\ref{lim3}) of the PDF for
velocity differences is actually a reflection of the presence of many
correlation lengths in the problem, which in turn is a consequence of the
anomalous scaling of correlation functions and, ultimately, the
intermittency of the velocity field in Burgers turbulence.

Let $n>>1$ be a large even positive integer. We know from (\ref{sn}) that as 
$\yb$ approaches zero 
\[
S_{n} (L(t)\yb,t) \approx \frac{1}{\sqrt{\pi}} 
\Gamma \bigg( \frac{n+2}{4} \bigg) \frac{L(t)}{L} 
U^{n} (t) \yb.
\]
For large $\yb$ however one expects the quantities 
$u(L(t)\yb ,t)$ and $u(0,t)$
to become independent. When this happens we have 
\begin{eqnarray*}
S_n(L(t)\yb ,t) & = & \left\langle (u(L(t)y,t) - u(0,t))^n \right\rangle \\
& \sim & \left\langle u^n(L(t)y,t) \right\rangle + \left\langle u^n(0,t) \right\rangle \\
& \sim & 2M_n = 2\frac{\Gamma (\frac{n+3}{4})}{\sqrt{\pi} (n+1)} \frac{L(t)}{L%
} U(t)^{n}.
\end{eqnarray*}
Here the cross terms in expanding the nth power are, using the independence,
of order $O(L^{-2})$. The region in between these two formulae for large and
small $y$ marks the correlation length for the nth moments. If we assume
there is a simple crossover then we can locate the scale at which it occurs
by equating the expressions for large $\yb$ 
and small $\yb$. These become equal,
i.e. $S_n(L(t)\yb ,t) \approx 2M_n$, at the value $n^{-3/4}$ and so the
correlation length for the n-th structure function is 
\begin{eqnarray}
L_{n} \sim \frac{L(t)}{n^{3/4}}, \quad n>>1  \label{ln}
\end{eqnarray}
and this shows the presence of many scales in our problem.

To show how this multiscaling is related to the crossover between the
asymptotic regimes (\ref{lim2}) and (\ref{lim3}) we shall use the PDF for
velocity differences to compute $S_{n} (y,t)$ for $n$ positive and large.
Writing $S_n(L(t)\yb ,t)$ as an integral against the PDF of 
$\Delta u(L(t)\yb ,t)$
and treating $n$ as a large parameter we see that the integral is dominated
by values of $|\ub|$ coming from the neighbourhood of the negative critical
point of the function 
\begin{eqnarray*}
F(u) = |\ub|^{n} \exp(-|\ub|^4)
\end{eqnarray*}
namely near $\ub _{c} = - n^{1/4}$. 
Note this value is much less than $-1$ for $%
n>>1$ and so we may neglect the part of the integral that uses the PDF in
the form (\ref{lim1}) and also neglect positive values of $\Delta u(y)$.
Now, if in addition $|\ub _{c}| << \yb ^{-1/3}$, 
we have to use asymptotics (\ref
{lim2}) to evaluate the contribution from the 
critical point, which yields $%
S_{n} (L(t)\yb ,t) \approx C\yb $. 
If $| \ub _{c} | >> \yb ^{-1/3}$ we have to use
asymptotics (\ref{lim3}) in our computations, which gives $S_{n} (L(t)\yb ,t)
\approx \mbox{Constant} $.  The crossover between these two answers
corresponds to the crossover between the asymptotics (\ref{lim1}) and (\ref
{lim2}) and occurs when $\yb = | \ub _{c}|^3 = n^{-3/4}$, exactly as in our
computed correlation length $L_n$ for the $n$-th structure function.

It remains to remark that multiscaling, and consequently a PDF for velocity
differences which has a crossover between a regime scaling like $y$ and one
that is independent of $y$, should be a general feature of DBT regardless of
the initial distribution. All related questions concerning other statistics
can be studied in more general situations, if one assumes a one-shock
approximation is valid, by using the information about the tails of the
one-point PDF of shock strength obtained in $\cite{sinai},
\cite{av2},\cite{av3}$.

It is worth noting that the presence of the multitude of correlation lengths
in Burgers turbulence was understood long ago by Robert Kraichnan, \cite
{Kraichnanms}, and rediscovered within the instanton approach to the forced
Burgers turbulence, \cite{Grishams}. It is also worth stressing that in
models of chaotic systems which do not account for the effects of
intermittency, there is always a single universal correlation length. A good
example is served by random matrix models, see \cite{Mehta} for a review.

Finally, let us remark that if we define the integral scale as the scale of
scaling behaviour of correlation functions, we must immediately conclude
that there is no such unique scale, there is rather a family of them
parameterized by the order of correlation function. In other words, the
notion of the integral scale becomes $local$, and the notion of the
universal inertial range disappears. (See also \cite{FrischVerg} for the
general discussion about the multitude of $dissipative$ scales
based on a multifractal models). This should
be a general feature of all intermittent turbulent systems, for instance,
Navier-Stokes turbulence.

\section{Acknowledgements}

We are grateful to E. Balkovski, D. Elworthy, G. Falkovich, U. Frisch 
\footnote{%
Who asked the very useful, perhaps rhetorical, question "Why study white
noise Burgers turbulence at all?"}, J. Gibbon, K. Khanin, S. Kuksin, S.
Nazarenko, A. Newell, C. Vassilicos for the most illuminating discussions.
We are most grateful for the hospitality of the Department of Complex
Systems of Weizmann Institute of Science, where part of this work has
been carried out.
The financial support through the research grant MA1117 from the
University of Warwick is also greatly appreciated.

\section{Note added in proof.} We are grateful to the referee of our
paper who drew our attention to  a recent preprint by L. Frachebourg
and Ph. A. Martin, \cite{Martin}, in which the study of the
model of decaying Burgers turbulence initiated by white noise
initial conditions (without compactness assumption) has been 
effectively completed. This model was originally considered by
Burgers himself about forty years ago but complexity of analysis
prevented him from obtaining explicit answers for anything but
the two- and three-point correlation functions of velocity field.
Now most of the questions about the statistics of velocity field in Burgers'
model can be effectively resolved using the integral representation
of the Green's function of a diffusion equation in the $(x,t)$-domain
with parabolic boundary derived in the above mentioned paper. 

\section{Appendix.}

In order to bound the various error terms in section 3 we will need to bound
the size of the true solution $u$, the asymptotic solution $u^*$ and the
size of their supports (i.e. the interval on which they are non-zero). We
use details from the method of construction of the vanishing viscosity
solution as descibed in \cite{Hopf} and recalled in section 2.

Suppose that initial velocity profile is supported in the interval $%
(x_{0}-l,x_{0}+l)$. The rightmost (respectively leftmost) parabola in the
chain of parabolic arcs built on the initial potential will always lie to
the left (respectively right) of the parabola with the same curvature that
passes through the point $(x_{0}+l,-Q)$ (respectively $(x_{0}-l,-Q)$) and
assumes minimial value equal to $-P$ (respectively $0$). This immediately
implies that both $u$ and $u^*$ are supported in the interval $[y_*,y^*]$
where 
\begin{eqnarray}
y_{*} =x_{0} -l-\sqrt{-2tQ}, \quad y^{*} =x_{0} + l +\sqrt{2t(P-Q)}.
\label{sest}
\end{eqnarray}
Using the fact that both $u$ and $u^{*}$ vanish at the point within $%
[x_{0}-l,x_{0}+l]$ at which $q(x)$ achieves its global minimum, we also find
that $|u|$ and $|u^*|$ are bounded by $u_{\mbox{max}}$ where 
\begin{eqnarray}
u_{\mbox{max}} &=& \max \{ (y^{*}- (x_{0}-l))/ t, ((x_0+l) -y_{*})/t \} 
\nonumber \\
& = & \max \{ (2l+ \sqrt{2t(P-Q)})/t, (2l+\sqrt{-2tQ})/t \} .  \label{um}
\end{eqnarray}
Estimates  (\ref{sest}),  (\ref{um}) and the bound (\ref{ts})  will be used to estimate all 
relevant error terms in section 3. The careful analysis
of these error terms leads to a better understanding for when the asymptotics for 
various statistics start to hold. 
\subsection{Proof of the estimate (\ref{Rnest}).}
Applying the above  estimates to the error term (\ref{Rnest}) we obtain 
\begin{eqnarray*}
| R_{n}(t)| & = & \langle | u^{n} (0,t) - u^{* n} (0,t)| \theta (T^{*} - t)
\rangle \\
& = & \langle | u^{n} (0,t) - u^{* n} (0,t)| \chi _{[y_{*},y^{*}]} (0) \;
\theta (T^{*} - t) \rangle \\
& \leq & 2 \langle u_{\mbox{max}}^{n} \chi _{[y_{*},y^{*}]} (0) \; \theta
(T^{*} - t) \rangle \\
& \leq & \frac{2}{L} \langle (y^{*}-y_{*}) u_{\mbox{max}}^{n}
\theta(T^{*}-t) \rangle \quad \mbox{(averaging over $x_0$)} \\
& \leq & \frac{2}{L} \langle (y^{*}-y_{*})^2 u_{\mbox{max}}^{2n}
\rangle^{1/2} \langle\theta(T^{*}-t) \rangle^{1/2} \quad %
\mbox{(Cauchy-Schwartz)} \\
& \leq & C_{n} \frac{L(t)}{L} U^{n} (t) (\frac{t^{*}}{t})^{1/4},
\end{eqnarray*}
where the last inequality uses the the estimate (\ref{ts}) and an
explicit calculation using $\rho(p,q)$. Comparing the first and the last
entries of the presented chain of inequalities we obtain a proof of (\ref
{Rnest}).

\subsection{Proof of the estimate on $r_{n}(t)$ from section (\ref{3.1}).}

We may bound $r_{n}(t)$ as follows: 
\begin{eqnarray*}
|r_{n} (t) | & \leq & \int_{-l}^{l} dx^{*} \int dp dq \;  \rho (p, q, x^*)
(\int_{-L}^{-L+x^*} + \int_{L}^{L+x^*} ) \frac{dx_{0}}{2L} 
| U (x_0,0,t,p,q)|^n \\
& \leq &  \int dp dq \;  \rho (p, q)
(\int_{-L-l}^{-L+l} + \int_{L-l}^{L+l} ) \frac{dx_{0}}{2L} 
| U (x_0,0,t,p,q)|^n \\
& \leq & \frac{2l}{L} \left( \frac{L+l}{L}\right)^n  \int dp dq \;  \rho (p, q)
\theta (\sqrt{-2Qt} -(L-l)) U^{n}(t)\\
& \leq & \frac{2l}{L} \left( \frac{L+l}{L}\right)^n  
\left( \frac{L(t)}{L-l} \right)^2 \exp 
\left(- \left( \frac{L-l}{L(t)} \right)^4 \right) U^{n}(t)
\end{eqnarray*}
where the last inequality follows by an explicit calculation using $\rho(p,q)$.
This is exponentially small in $L$ and so does not affect the asymptotics which take the limit
$L \rightarrow \infty$ first and preserve only the $O(L^{-1})$ terms. A similar argument controls
similar error terms of this form for the other statistics considered.
\subsection{Proof of the estimate (\ref{Rest}).}
The proof of (\ref{Rest}) is  similar to that of (\ref{Rnest}):
\begin{eqnarray*}
| R(u,t)| &  \leq & 
= \langle | \theta(u (0,t)-u) -\theta (u^{*} (0,t)-u) | 
\chi_{[y_{*},y^{*}]} (0) \theta (T^{*} - t) \rangle \\
& \leq & 2 \langle \chi _{[y_{*},y^{*}]} (0) \theta (T^{*} - t) \rangle \\
& \leq & \frac{1}{L} \langle (y^{*}-y_{*}) \theta(T^{*}-t) \rangle \\
& \leq  & \frac{1}{L} \langle (y^{*}-y_{*})^2 \rangle^{1/2}
\langle\theta(T^{*}-t) \rangle^{1/2} \\
& \leq & C \frac{L(t)}{L} (\frac{t_{c}}{t})^{1/4}.
\end{eqnarray*}
\subsection{Proof of the estimate (\ref{est1}).}
We can split this error term into two via
\begin{equation} \label{est1a}
 |R_n(y,t)| \leq 2^n \langle \Delta u(y,t)^n \theta(T^*-t) \rangle
+ 2^n \langle \Delta u^*(y,t)^n \theta(T^*-t) \rangle ,
\end{equation}
where $\Delta u(y) = u(y/2,t)-u(-y/2,t)$.
We show how to bound the first of these terms, the other being
entirely similar. The vanishing viscosity solution $u$ takes the form, 
within its support, 
of a line with slope $1/t$ plus a series of downward jumps. So we may define
$F(x,t)$ to be a non increasing piecewise constant function so that, for
$x$ in the support of  $u$,  
\begin{eqnarray*}
u(y,t)= \frac{y-x_{0}}{t}+F(y-x_0,t).
\end{eqnarray*}
It is easy to see that $|\Delta F(y,t)| =
|F(y/2,t)-F(-y/2,t)| \leq 2 \umm$.
Also $|\Delta u(y,t)|  \leq |y/t| + |\Delta F(y-x_0,t)|$ whenever one of the points
$y/2$ or $-y/2$ is in the support of $u$. 
So we bound the first term on the right hand side of (\ref{est1a}) by
\begin{eqnarray}
&& \langle (|y/t|+ |\Delta F(y-x_0)|)^n \chi_{[y_* -(y/2), y^*+(y/2)]}(0) \theta(T^*-t) \rangle 
\nonumber \\
& \leq & 2^n  \langle  |\Delta F(y-x_0)|^n  \theta(T^*-t) \rangle 
+ 2^n  \langle |y/t|^n  \chi_{[y_* -(y/2), y^*+(y/2)]}(0) \theta(T^*-t) \rangle. 
\label{est1b}
\end{eqnarray}
The first term on the right hand side
of (\ref{est1b})  can be bounded  by averaging over $x_0$ first and using
\begin{eqnarray*}
 \int_{-L}^{L} \frac{dx_{0}}{2L} |\Delta F(y-x_0,t)|^n
&  \leq & (2 \umm)^{n-1}
\int_{-L}^{L} \frac{dx_{0}}{2L} |\Delta F(y-x_0,t)| \\
& \leq & (2 \umm)^{n-1} \frac{y \umm}{L},
\end{eqnarray*}
using in the last inequality the fact that $F$ is decreasing and bounded by 
$2\umm$.
Substituting into (\ref{est1b}) one can take the further averaging as for previous error bounds. 
By taking $t$ large enough that $L(t) \geq y$ and 
combining the various terms one 
arrives at the desired error bound. 

\subsection{The proof of the estimate (\ref{errtrm}).}

The proof of this estimate is similar to that of (\ref{est1}).
 Noting that $\Delta u(y,t) = \Delta F(y-x_0,t) + (y/t)$ we may
write
\begin{eqnarray*}
\int \frac{dx_0}{2L} \theta(u-\Delta u(y,t)) &=& \int \frac{dx_0}{2L} 
\theta( |\Delta F(y-x_0)| -(y/t)-|u|) \\
& \leq & \int \frac{dx_0}{2L} \frac{|\Delta F(y-x_0)|}{(y/t)+|u|} \\
& \leq & \frac{1}{2L} \frac{2y \umm}{(y/t)+|u|)}.
\end{eqnarray*}
A similar estimate holds for $\Delta u^*(y,t)$. Hence
\begin{eqnarray*}
| R(u,y,t)| & \leq & \langle \left(\theta(u-\Delta u(y,t)) + \theta(u-\Delta u^*(y,t) \right) 
\theta (T^*-t) \rangle \\
& \leq &  \frac{1}{L} \frac{2y}{(y/t)+|u|)} \langle \umm \theta(T^*-t) \rangle \\
& \leq & \frac{L(t)}{L} \frac{\bar{y}}{\bar{y}+|\bar{u}|} \left( \frac{t_c}{t} \right)^{1/4}.
\end{eqnarray*}
\subsection{Proof of the estimate (\ref{rnt}).}

The proof of this estimate is similar to that of (\ref{errtrm}). The key
change is to obtain a bound for
\begin{equation} \label{timeerror}
\int \frac{dx_0}{2L} | F(y-x_0,t) - F(y-x_0,t+\tau)|.
\end{equation}
The piecewise constant profile $F(y,t)$ consists of a series
of shocks which may travel forwards or backwards but move with a
maximum speed $\umm$. The total height of the shocks is also
bounded by $\umm$. So the integral (\ref{timeerror}) can be
bounded by $\umm^2 \tau/2L$. The possibility of infinitely many
shocks, or the merging of shocks between times $t$ and $t+\tau$,
does not affect this upper bound. 

\subsection{The proof of the estimate (\ref{ts}).}
The construction of the two shock
profile uses two parabolas that pass through the graph of the Brownian
motion $-q(x) = -\int^x_{-\infty} u_0(z)dz$ at its point of maximum. Below is a lemma
about the behavior of a Brownian path near its maximum.
\begin{Lemma}
\label{appendixlemma} Let $(B_t: 0 \leq t \leq 1)$ be a standard Brownian
motion started at zero. Define 
\[
M= \sup_{t \in [0,1]} B_t, \quad \Sigma = \inf \{t:B_t=M\}. 
\]
We consider the pieces of the path $(B_t)$ either side of its maximum by
defining 
\[
X_t = M- B_{\Sigma-t} \mbox{ for $t \in [0,\Sigma]$}, \quad \bar{X}_t = M-
B_{\Sigma+t} \mbox{ for $t \in [0,1-\Sigma]$}. 
\]
Define the slopes of two lines that pass through the maximum and lie
above the path by 
\[
\Theta = \inf \{ X_t/t: 0 < t \leq \Sigma\}, \quad \bar{\Theta} = \inf \{ 
\bar{X}_t/t: 0 < t \leq 1-\Sigma\}. 
\]

\begin{enumerate}
\item[a)]  The triples $(M ,\Sigma, (X_t:t \leq \Sigma))$ and $(M-B_1,
1-\Sigma, (\bar{X}_t:t \leq 1-\Sigma))$ are identically distributed.

\item[b)]  The law of $(M,\Sigma)$ is given by 
\[
P(M \in dm, \Sigma \in d \sigma) = \frac{m \sigma^{-1}}{ \pi
(\sigma(1-\sigma))^{1/2}} \exp (-m^2/2 \sigma) dm \, d\sigma. 
\]

\item[c)]  Conditional on $M \in dm, \Sigma \in d\sigma $ the path $(X_t: t
\leq \sigma)$ satisfies $X_0=0$ and solves the stochastic differential
equation, driven by a Brownian motion $(W_t)$, 
\begin{equation}  \label{SDE}
dX_t = f(t,X_t)dt + dW_t, \quad 
\mbox{where $f(t,x)= \frac{m-x}{\sigma-t} + \frac{2m}{\sigma-t}
(\exp ( \frac{2mx}{\sigma-t}) -1)^{-1}$.}
\end{equation}

\item[d)]  For $(X_t)$ that solves (\ref{SDE}) we have the estimate 
\[
P(\Theta \leq \theta) \leq C \theta (m + \sigma m^{-1}) + I(m \leq \theta
\sigma) . 
\]
\end{enumerate}
\end{Lemma}

We delay the proof of this lemma until the end of this appendix and first
use it to prove the estimate (\ref{ts}) on the tail $P(T^* \geq t)$ of the time $T^*$
at which the two shock profile is obtained. The construction of the two
shock profile uses the function $q(x)= \int^x_{\infty} u_0(z)dz$, its global
minimum $Q$ and the position $x_0 + x^*$ at which the minimum is attained.
Two parabolas of the form $\pi(z)= (z-x)^2/2t$ (and $\bar{\pi}(z)= P+ (z-%
\bar{x})^2/2t$) are constructed to pass through the point $(x_0 +x^*,-Q)$.
The slopes of the parabolas at the point $x_0+x^*$ are $(-2Q/t)^{1/2}$
(respectively $(2(P-Q)/t)^{1/2}$). Let $T$ (respectively $\bar{T}$) be the
smallest time $t$ at which the parabola $\pi$ (respectively $\bar{\pi}$)
lies above the graph of $-q(x)$. Then the two shock profile is attained for
times $t\geq T^*=\max \{T,\bar{T}\}$.

To apply the lemma we must rescale to obtain a standard Brownian path of length one.
Set $B_t = -(2lJ)^{-1/2} q(x_0-l+2lt)$ for $t \in [0,1]$. Then $(B_t)$ is a
standard Brownian motion and its maximum $M $ takes the value $-Q
/(2lJ)^{1/2}$. The construction of the parabola $\pi$ (respectively $\bar{\pi%
}$) show that if $\Theta \leq (-4lQ/tJ)^{1/2}$ then $t \leq T$ (respectively
if $\bar{\Theta} \leq ((4l(P-Q)/tJ)^{1/2} $ then $t \leq \bar{T}$). Part a)
of the lemma shows that both of these events have the same probability. So,
applying part d) of the lemma, 
\begin{eqnarray*}
P(T^* \geq t) & \leq & 2 P(\Theta \leq (-4lQ/tJ)^{1/2}) \\
&=& 2 P(\Theta \leq 2^{5/4} M^{1/2} t^{-1/2} l^{3/4}J^{-1/4} ) \\
& \leq & C t^{-1/2} l^{3/4} J^{-1/4} E(M^{1/2} (M + \Sigma M^{-1}) ) \\ 
+2P (M \leq 2^{5/4} M^{1/2} t^{-1/2} l^{3/4}J^{-1/4} \Sigma) \\
& \leq & C t^{-1/2} l^{3/4}J^{-1/4}
\end{eqnarray*}
using Markov's inequality in the last inequality and the exact distribution
of $(M,\Sigma)$ in part b) of the lemma. This completes the proof of (\ref{ts})
and it remains to describe the proof of the lemma.

Part a) of the lemma follows from the symmetry of the problem with respect
to the time reversal $t \rightarrow 1 -t$. The distribution of $(M,\Sigma)$
is well known and may be obtained for example by exploiting the reflection
principle. Conditional on $M \in dm, \Sigma \in \sigma$ the path $(X_t)$
becomes a Brownian bridge, taking the value zero at time zero and the value $%
m$ at time $\sigma$, that is conditioned to never take negative values. The
equation describing the evolution can then be obtained using an h-transform
as in Rogers and Williams \cite{Rogers} section 4.23.

We first sketch the idea for estimating $P(\Theta \leq \theta )= P(X_s
<\theta s \; \mbox{for some $s \leq \sigma$}).$ The drift $f(t,x)$ in
equation (\ref{SDE}) is approximately $1/x$ for small $t$ and $x$. If this
approximation were exact the process $(X_t)$ would satisfy $dX= X^{-1} dt +
dW$ which is uniquely solved by the three dimensional Bessel process (the
radius of a three dimensional Brownian motion). For a Bessel process one can
make use of time inversion via the identity in distribution 
\[
(X_t:t>0) = (tX_{1/t}:t>0) 
\]
and potential theory for three dimensional Brownian motion which gives 
\[
P(X_s < \theta \; \mbox{for some $s \geq 0$}| X_0 =x) = \min \{\theta
x^{-1}, 1 \}. 
\]
Then 
\begin{eqnarray*}
P(X_s <\theta s \; \mbox{for some $s \leq \sigma$}) & = & P(X_s < \theta \; %
\mbox{for some $s \geq 1/\sigma$}) \\
&=& E( \min \{ \theta X_{1/\sigma}^{-1} ,1\}) \\
&=& E( \min \{ \theta \sigma^{1/2} X_{1}^{-1}, 1\}) \\
&=& \int_0^{\infty} (2 \pi)^{-3/2} r^2 \exp(-r^2/2) \min \{ \theta
\sigma^{1/2} r^{-1},1\} dr \\
& \leq & C \theta \sigma^{1/2}
\end{eqnarray*}
where the penultimate equality follows from Brownian scaling and the final
equality from a calculation using the density of the Gaussian variable $X_1$%
. To exploit this idea we divide the interval $[0,\sigma]$ into two parts,
over the first of which the approximation $f(t,x) \approx 1/x$ is
sufficiently good.

We first estimate $P(X_s <\theta s \; \mbox{for some $s \leq \sigma/2 $})$.
Using the elementary inequalities $(1-z)/2z \leq (e^{2z}-1)^{-1} \leq 1/2z $
for all $z>0$ one obtains the bounds $x^{-1} - x(\sigma -t)^{-1} \leq f(t,x)
\leq x^{-1} + 2m \sigma^{-1}$. Hence 
\begin{equation}  \label{appendix44}
X_t^{-1} -2\sigma^{-1} X_t \leq f(t,X_t) \leq X_t^{-1} + 2 m (\sigma-t)^{-1}
\qquad \mbox{for $t \leq \sigma/2 $}.
\end{equation}
So the solution of the equation 
\[
dY_t = Y_t^{-1} dt - 2 \sigma^{-1} Y_t dt + dW_t, \;\; Y_0=0 
\]
satisfies $Y_t \leq X_t$ for all $t \leq \tau $. To remove the unwanted $-2
\sigma^{-1} Y_t dt$ in the drift of $(Y_t)$ we use a change of measure.
Define a new probability measure $Q$ by defining the Radon-Nicodym
derivative $M$ by 
\begin{eqnarray*}
M &=& \left. \frac{dQ}{dP} \right|_{\mathcal{F}_{\sigma/2} } \\
& = & \exp( 2 \sigma^{-1} \int^{\sigma/2}_0 Y_s dW_s- 2 \sigma^{-2}
\int^{\sigma/2}_0 Y_s^2 ds) \\
& = & \exp( \sigma^{-1} Y_{\sigma/2}^2 + 2 \sigma^{-2} \int^{\sigma/2}_0
Y^2_s ds - 3/2) \\
& \geq & \exp(-3/2).
\end{eqnarray*}
The second equality here follows from Ito's formula. By Girsanov's theorem ( see
\cite{Revuz}) the process $(Y_t)$ solves $dY=Y^{-1} dt + d\tilde{W}$ with
respect to some Brownian motion $(\tilde{W})$ under $Q$, implying that $(Y_t)
$ is a three dimensional Bessel process under $Q$. Writing $E_Q$ for the
expectation under $Q$ we have 
\begin{eqnarray}
P(X_s <\theta s \; \mbox{for some $s \leq \sigma/2  $}) & \leq & P(Y_s
<\theta s \; \mbox{for some $s \leq \sigma/2 $})  \nonumber \\
& = & E_Q ( M^{-1} I(Y_s <\theta s \; \mbox{for some $s \leq \sigma/2 $})) 
\nonumber \\
& \leq & e^{3/2} Q(Y_s <\theta s \; \mbox{for some $s \leq \sigma/2 $}) 
\nonumber \\
& \leq & C \theta \sigma^{1/2}  \label{appendix25}
\end{eqnarray}
using the argument given above.

It remains to estimate the probability $P(X_s <\theta s \; 
\mbox{for some
$\sigma/2 \leq s \leq \sigma $})$. We shall further condition on the value
of $X_{\sigma/2}$. If $X_{\sigma/2} \in dr$ the evolution of $(X_s:s \in
[\sigma/2,\sigma])$ is that of a Brownian bridge starting at r, ending at m
and conditioned to take non-negative values. We write $Q_x$ for the law of a
one-dimensional Brownian motion $(W_t)$ started at $x$ and we define $H_a=
\inf \{t: W_t \leq a\}$. Then, supposing $r,q \geq \theta \sigma$, we have 
\begin{eqnarray*}
&& P(X_s \leq \theta s \; \mbox{for some $s \in [\sigma/2, \sigma]$}|
X_{\sigma/2} \in dr) \\
& = & 1- P(X_s >\theta s \; \mbox{for all $s \in [\sigma/2, \sigma]$}|
X_{\sigma/2} \in dr) \\
& \leq & 1- P(X_s >\theta \sigma \; \mbox{for all $s \in [\sigma/2, \sigma]$}%
| X_{\sigma/2} \in dr) \\
& = & 1- Q_r (H_{\theta \sigma} > \sigma/2| W_{\sigma/2} \in dm, H_0 >
\sigma/2) \\
& = & 1- \frac{Q_r (H_{\theta \sigma} > \sigma/2, W_{\sigma/2} \in dm)}{Q_r
(H_0 > \sigma/2, W_{\sigma/2} \in dm)}
\end{eqnarray*}
The reflection principle can be used to show that, for $%
a \leq r,m$, 
\begin{equation}
Q_r ( H_a > t, W_t \in dm) = \left( p_t(m-r) - p_t(m+r-2a) \right) dm
\end{equation}
where $p_t(z) = (2 \pi t)^{-1/2} \exp (-z^2/2t)$. Using this we rewrite the
last expression as 
\begin{eqnarray*}
& & \frac{ \exp ( (m+r )^2/\sigma) - \exp ( (m+r-2 \theta \sigma)^2/\sigma) 
}{\exp ( (m+r)^2/\sigma) - \exp ( (m-r)^2/\sigma) } \\
& = & (1- \exp( \frac{-4mr}{\sigma}))^{-1} 4 \theta (m+r-2\eta) \exp ( \frac{%
(m+r-2\eta)^2 -(m+r)^2}{\sigma} ) \\
&& \hspace{1in} 
\mbox{ for some $\eta \in [0, \theta \sigma]$ by the mean
value theorem} \\
& \leq & (1- \exp( \frac{-4mr}{\sigma}))^{-1} 4 \theta (m+r-2\eta) \\
& \leq & C \theta (1+ \frac{\sigma}{4mr})(m+r) \qquad 
\mbox{(using 
               $(1-e^{-z})^{-1} \leq C(1+z^{-1})$)} \\
& \leq & C \theta (m+r + \sigma r^{-1} + \sigma m^{-1}) .
\end{eqnarray*}
Thus 
\begin{eqnarray}  
&& P(X_s \leq \theta s \; \mbox{for some $s \in [\tau, \sigma]$}| X_{\tau}
\in dr)  \nonumber \\
& \leq & C \theta (m+r + \sigma r^{-1} + \sigma m^{-1}) + I(r \leq \theta
\sigma) + I(m \leq \theta \sigma).  \label{appendix30}
\end{eqnarray}
We now undo the conditioning on $X_{\sigma/2} \in dr$. Using the upper bound
in (\ref{appendix44}) and Ito's formula one obtains $dX^2_t \leq
(3+4m\sigma^{-1}X_t)dt + 2X_t dW_t$. Taking expectations one has 
\begin{eqnarray*}
E(X_t^2) & \leq & 3t + 4m \sigma^{-1} \int^t_0 E(X_s) ds \\
& \leq & (3 + m^2 \sigma^{-1})t + 4 \sigma^{-1} \int^t_0 E(X_s^2) ds.
\end{eqnarray*}
Applying Gronwall's inequality shows that $E(X_{\sigma/2}) \leq
(E(X_{\sigma/2}^2))^{1/2} \leq C (\sigma^{1/2} +m)$. By Markov's inequality $%
P(X_{\tau} \leq \theta \sigma) \leq \theta \sigma E(X_{\tau}^{-1})$. Using
the comparison with a Bessel process as before we have $E(X_{\tau}^{-1})
\leq e^{3/2} E_Q(Y_{\sigma/2 }^{-1} ) \leq C \sigma^{-1/2}.$ Using these
bounds in (\ref{appendix30}) and combining with (\ref{appendix25}) leads to
the estimate in part d) of the lemma.


\begin{thebibliography}{99}

\bibitem{aurell}  E. Aurell, U. Frisch, A. Noullez, M. Blank \emph{%
Bifractality of the Devil's staircase appearing in the Burgers equation with
Brownian initial velocity}, chao-dyn/961101, published in
{\it J. Stat. Phys. \bf 88}, 1151--1164;
\bibitem{av1}  Avellaneda, M.; Ryan, R.; E, Weinan \emph{PDFs for velocity
and velocity gradients in Burgers' turbulence}, Comm. Math. Phys. 172 (1995),
no. 1, 13--38;

\bibitem{av2}  Avellaneda, Marco; E, Weinan \emph{Statistical properties of
shocks in Burgers turbulence.} Comm. Math. Phys. 172 (1995), no. 1, 13--38;

\bibitem{av3}  Avellaneda, Marco \emph{Statistical properties of shocks in
Burgers turbulence. II. Tail probabilities for velocities, shock-strengths
and rarefaction intervals.} Comm. Math. Phys. 169 (1995), no. 1, 45--59;

\bibitem{grisha}  E. Balkovsky, G. Falkovich, I. Kolokolov, V. Lebedev 
\emph{Intermittency of Burgers' Turbulence }, Phys. Rev. Lett, 78, 1452,
(1997);

\bibitem{grisha1}  E. Balkovsky, G. Falkovich, I. Kolokolov, V. Lebedev 
\emph{Viscous Instanton for Burgers' Turbulence}, Phys. Rev. Lett., 78,
1452, (1997);


\bibitem{Grisha}  E. Balkovski and G. Falkovich {\em Private communication} ;

\bibitem{FrischBec}  J. Bec, U. Frisch \emph{Pdf's of Derivatives and
Increments for Decaying Burgers Turbulence}, cond-mat/9906047;

\bibitem{Gaw}  D. Bernard and K. Gawedzki \emph{Scaling and exotic regimes
in the decaying Burgers turbulence}, chao-dyn/9805002;

\bibitem{Borodin}  Borodin, Andrei N.; Salminen, Paavo 
{\em Handbook of Brownian motion---facts and formulae.}
Probability and its Applications. 
Birkhäuser Verlag, Basel, 1996. xiv+462 pp. ISBN: 3-7643-5463-1;

\bibitem{Burgers}  Burgers, J. M. \emph{Statistical problems connected with
the solution of a simple non-linear partial differential equation.} I, II,
III. Nederl. Akad. Wetensch. Proc. Ser. B. \textbf{57} (1954),403--413,
414--424, 425--433;

\bibitem{Grishams} G. Falkovich {\em Unpublished};

\bibitem{Martin} L. Frachebourg, Ph. A. Martin,
{\em Exact statistical properties of the Burgers equation},
cond-mat/9909056;

\bibitem{FrischVerg} Frisch, U.; Vergassola, M. 
{\em A prediction of the multifractal model: the intermediate dissipation
range.} 
New approaches and concepts in turbulence (Monte Verità, 1991), 
29--34, Monte Verità, Birkhäuser, Basel, 1993;

\bibitem{Frisch}  Frisch, Uriel \emph{Turbulence. The legacy of A. N.
Kolmogorov.} Cambridge University Press, Cambridge, 1995. xiv+296 pp. ISBN:;


\bibitem{bob}  T. Gotoh and R. Kraichnan \emph{Statistics of decaying
Burgers turbulence}, Phys. Fluids A \textbf{5} (2) (1993);

\bibitem{Gurbatov}  Gurbatov, S. N.; Malakhov, A. N.; Saichev, A. I. \emph{%
Nonlinear random waves and turbulence in nondispersive media: waves, rays,
particles}. Translated from the Russian. Supplement 1 by Adrian L. Melott
and Sergei F. Shandarin. Supplement 2 by V. I. Arnol'd, Yu. M. Baryshnikov
and I. A. Bogayevsky. Translation edited and with a preface by D. G.
Crighton. Nonlinear Science: Theory and Applications. Manchester University
Press, Manchester, 1991. x+308 pp.;

\bibitem{Gurbatov1}  Gurbatov, S. N.; Simdyankin, S. I.; Aurell, E.; Frisch,
U.; T\'{o}th, G. \emph{On the decay of Burgers turbulence.} J. Fluid Mech.
344 (1997), 339--374;

\bibitem{migdal}  V. Gurarie, A. Migdal \emph{Instantons in Burgers
Equation,} Phys.Rev. E54 (1996) 4908-4914;

\bibitem{Hopf}  E. Hopf {\em The partial differential equation $u_{t} + uu_{x}
= \mu u_{xx}$}; Comm. Pure Appl. Math. \textbf{3}, (1950). 201--230;

\bibitem{khanin}  E, W., Khanin, K., Mazel A., and Sinai Ya. G. \emph{%
Invariant measures for the random forced Burgers equation}, submitted to
Ann. Math.;

\bibitem{khanin1}  E, W., Khanin, K., Mazel A., and Sinai Ya. G. \emph{%
Probability distribution functions for the random forced Burgers equation},
Phys. Rev. Letters \textbf{78}, 1904-1907 (1997);

\bibitem{Kida}  Kida, Shigeo \emph{Asymptotic properties of Burgers
turbulence.} J. Fluid Mech. 93 (1979), no. 2, 337--377;

\bibitem{bob1}  Robert H. Kraichnan \emph{Note on Forced Burgers Turbulence}%
, chao-dyn/9901023;

\bibitem{Kraichnanms} R. Kraichnan {\em Unpublished};

\bibitem{Mehta} Mehta, Madan Lal 
{\em Random matrices.} Second edition. Academic Press, Inc., Boston, MA, 1991.
xviii+562 pp. ISBN: 0-12-488051-7;

\bibitem{parker}  Parker, D. F. \emph{The decay of sawtooth solutions to the
Burgers equation}, Proc. Roy. Soc. London Ser. A 369 (1980), no. 1738,
409--424;

\bibitem{Polyakov}  Polyakov, A. M. \emph{Turbulence without pressure.}
Phys. Rev. E (3) 52 (1995), no. 6, part A, 6183--6188;

\bibitem{Revuz}  Revuz, D. and Yor, M. \emph{Continuous Martingales and
Brownian Motion}, Springer verlag, 1991;

\bibitem{Rogers} Rogers, L.C.G. and Williams, D. \emph{Diffusions, Markov
Processes, and Martingales}. Volume 2. Wiley,  1986;

\bibitem{av}  Ryan, Reade; Avellaneda, Marco \emph{The one-point statistics
of viscous Burgers turbulence initialized with Gaussian data}, Comm. Math.
Phys. 200 (1999), no. 1, 1--23;

\bibitem{sinai}  Sinai, Ya. G. \emph{Statistics of shocks in solutions of
inviscid Burgers equation}, Comm. Math. Phys. 148 (1992), no. 3, 601--621;

\bibitem{truman}  Truman, A.; Zhao, H. Z. \emph{On stochastic diffusion
equations and stochastic Burgers' equations.} J. Math. Phys. 37 (1996), no.
1, 283--307;

\bibitem{Truman}  Truman, A.; Zhao, H. Z. \emph{Stochastic Burgers'
equations and their semi-classical expansions}, Comm. Math. Phys. 194
(1998), \textbf{1}, 231--248;

\bibitem{weinan}  Weinan E, Eric Vanden Eijnden, \emph{Statistical Theory
for the Stochastic Burgers Equation in the Inviscid Limit}, chao-dyn/9904028.

\end{thebibliography}
\end{document}